\documentclass[prc,aps,eqsecnum,epsf,amssymb]{revtex4}
\usepackage{graphicx}

\newcommand{\bmp}{{\mbox{\boldmath $p$}}}

\newcommand{\bmq}{{\mbox{\boldmath $q$}}}

\newcommand{\qq}{{|\mbox{\boldmath $q$}|}}

\begin{document}
\preprint {WIS-02/26 Dec-DPP}
\date{\today}
\title{The neutron magnetic form factor $G_M^n(Q^2)$ from Quasi-Elastic
inclusive scattering data on $D$ and $^4$He}
\author{A.S. Rinat and M.F. Taragin}
\address{Weizmann Institute of Science, Department of Particle Physics,
Rehovot 76100, Israel}
\author{M. Viviani}
\address{INFN, Sezione Pisa and Phys. Dept., University of Pisa, I-56100,
Italy}
\begin{abstract}

We analyze cross sections for Quasi-Elastic inclusive scattering of 
electrons on nuclei and show that the observed isolated peaks
for relatively low $Q^2$ are unique for the lightest targets. 
Focusing in particular on  D and $^4$He, we investigate in two 
ways to what measure the above peaks can be allocated to 
nucleon-elastic processes. We first compute approximate upper limits 
for the nucleon-inelastic background in the Quasi-Elastic region 
due to inclusive $\Delta$ excitation, and find those to be small. Far 
more precise is a semi-phenomenological approach, where the dominance of 
nucleon-elastic processes is translated into a set of stringent 
requirements. We show that  
those are very well fulfilled for recent D data, and to a somewhat lesser 
extent for older D and $^4$He data. With knowledge of $G_{E,M}^p$ and 
information on $G_E^n$, we  then extract $G_M^n$ and find agreement with
values obtained by alternative methods. We discuss the sensitivity of the 
extraction method and mention future applications.

\end{abstract}

\maketitle

\section{Introduction.}

Charge-current distributions of  hadrons
are basic sources of information, which may be compared
with predictions of fundamental theories. Examples are
static form factors of the neutron and its structure functions (SF)
which depend on those distributions. 
Over many years, experimental efforts have been made 
to extract those observables with maximal accuracy. This requires 
high-quality data, and in parallel, accurate 
control of nuclear medium effects. In this note we focus on the
magnetic form factor of the neutron. 

A standard tool for the study of $G_M^n(Q^2)$ has been Quasi-Elastic 
(QE) electron scattering on a D for relatively low $Q^2$. We also
mention semi-inclusive scattering experiments D$(e,e'N)X\,$,  
where $N=p$ or $n$~\cite{ank1,kubon}, as well as total inclusive data
on D up to $Q^2\le4\,{\rm GeV}^2\,\,$ \cite{lung}. The varied kinematics 
in the latter experiment made it possible to perform  a Rosenbluth 
separation and a subsequent isolation of transverse parts ${\cal R}_T$ 
of cross sections. Once inelastic background effects 
are removed, one is left with a simple 
expression for ${\cal R}_T^{NE}\propto [(G_M^p)^2+(G_M^n)^2]$. 

Another source of information is the asymmetry in the inclusive process 
$\vec{{^3}{\rm He}}({\vec e},e')X\,\,$ \cite{gao,xu}, which requires 
for its analysis a complete 3-body  calculation. In  most of those one has
neglected Final State Interactions (FSI) or relativistic kinematics 
\cite{xu,golak}. The present range $Q^2\lesssim 
0.6\,$  GeV$^2$ will soon be considerably enlarged \cite{jones2}.

In the following we re-open the discussion on the extraction of $G_M^n$
from QE inclusive scattering on D and other targets. There is no change in 
the basic understanding of those reactions. New is the much  improved 
accuracy, for instance, with which one nowadays computes 
wave functions for light targets \cite{bench}. In parallel,
more precise expressions for FSI have been also obtained. The above new 
input is here applied to analyze the total-inclusive 
data for light nuclei.  

We base our analysis on a specific relation between nuclear and nucleon
structure functions. The latter leads to the definition of the Nucleon 
Elastic (NE) and Nucleon Inelastic (NI) components of the inclusive cross 
sections for a composite target, which correspond to processes where 
a virtual photon leaves a struck $N$ in its ground state, or excites it. 

In our analysis we consider recent D data \cite{arrd,nicu}, 
as well as older ones on $^4$He$\,$ \cite{day} and D \cite{lung}.
We first address inelastic contributions in the QE region. We estimate 
their magnitude on a model of inclusive $N$-$\Delta$ excitation and show
that those are small compared to the QE total inclusive cross section.
Next we formulate in a semi-empirical fashion stringent requirements which
have to be fulfilled if total inclusive cross sections are dominated by
their NE components. We find that those demands are accurately fulfilled 
for the recent D \cite{arrd,nicu} and to a somewhat lesser extent for 
the NE3 $^4$He$\,$ data \cite{day}.  
In the same fashion we re-analyze separated transverse parts of the 
above-mentioned older D data \cite{lung} and in parallel exploit the 
simultaneously measured total QE inclusive  cross sections, which before 
have not been investigated in their own right.

In the above NE parts appear all four static form factors 
$G_{E,M}^{p,n}(Q^2)$. Those  for a proton have recently been determined 
with improved precision \cite{sill,brash,mjon}, while 
$G_E^n$ is reasonably well known for $Q^2\le 1.6\,$GeV$^2\,\,$~\cite{rocco}. 
As a consequence one can extract $G_M^n$ from cross sections, provided 
those are indeed dominated by their NE components.

We show that the thus determined $G_M^n$ are essentially independent of
both the QE data points chosen for extraction, and of the target 
nucleus. We discuss the sensitivity of our results to the quality of the 
experimental input and mention forthcoming precise data to which
the presented extraction methods can be applied. Those will help to 
sharpen the results obtained below.

\section{Quasi-Elastic inclusive scattering.}

\subsection{Generalities.}\label{sec:gnr}

Consider the cross section per nucleon for inclusive scattering over 
an angle $\theta$ of unpolarized electrons, with initial and final beam 
energies $E,E-\nu$. The same, relative to the Mott cross 
section is 
\begin{eqnarray} 
  K^A(x,Q^2)&\equiv& 
  \frac{d^2\sigma^A(E;\theta,\nu)/A}{d\Omega\,d\nu} \bigg /
\sigma_M(E;\theta,\nu)
\nonumber\\
&=&\bigg\lbrack\frac {2xM}{Q^2}
  F_2^A(x,Q^2)+ \frac{2}{M}F_1^A(x,Q^2){\rm tan}^2(\theta/2) \bigg\rbrack
\label{a1}  
\end{eqnarray}
$F_{1,2}^A(x,Q^2)$ are the nuclear SF which depend on the modulus of the 
squared 4-momentum transfer $q^2=-Q^2=-(\qq^2-\nu^2)$ and on 
the Bjorken variable $x=Q^2/2M\nu$. With $M$ the nucleon mass, 
its range is $0\le x\le A$. 
In order to calculate the nuclear SF, we shall exploit a previously
postulated relation between nucleon and nuclear SF $\,$\cite{gr}, which for 
isospin $I=0$ targets reads 
\begin{eqnarray}                                                 
  F^A_k(x,Q^2)&=&\int_x^A\frac {dz}{z^{2-k}} f^{PN,A}(z,Q^2)\sum_l 
   C_{kl}(z,Q^2)
  \bigg [F_l^p \bigg (\frac {x}{z},Q^2\bigg )
  +F_l^n\bigg (\frac {x}{z},Q^2\bigg ) \bigg ]\bigg /2                
\label{a2}                                                    
\end{eqnarray}                                    
The link between the SF $F_{1,2}^A$ and the nucleon SF $F_{1,2}^{N=p,n}$  
(assumed to coincide with the free ones), is provided by the SF $f^{PN,A}$ 
of a fictitious target composed of $A$ point-nucleons. It includes the 
effect of the mixing of the nucleon SF 
via the coefficients $C_{kl}$, which expression can be obtained using 
standard procedures~\cite{atw,sss}. As usual, we take 
$C_{11}=1,\,C_{12}=C_{21}=0$, and retain only $C_{22}$ in the expression 
above.  In Appendix A we provide details. 

Many data-analyses have been made with $f^{PN,A}$, calculated in the 
Plane Wave Impulse Approximation (PWIA) in terms of the single-hole 
spectral function~\cite{ciof}. We favor the Gersch-Rodriguez-Smith (GRS) 
theory for $f^{PN,A}\,\,$ \cite{grs}, which has recently  been generalized 
for use in the relativistic regime \cite{gr1}. One of the reasons of our 
preference is the convergence of the GRS series to the exact $f^{PN,A}$, which 
is generally faster than is the case for the Impulse Series (IS). Moreover, 
it is more convenient to use the GRS series for a computation of FSI, 
which are present in $f^{PN,A}$ \cite{gr,gr1,rt}.

In the following we shall focus on  the immediate neighborhood  of the
Quasi-Elastic-Peak (QEP), $|x|\approx 1$, where nucleons, as described 
by Eq. (\ref{a2}), are the dominant parton sources (see for instance Ref.
\onlinecite{lle}). 

\subsection{Nucleon-Elastic and Inelastic components of SF.}\label{sec:NEI}

We first consider in Eq. (\ref{a2}) the SF $F_k^{N}$ of nucleons and 
separate those in nucleon-elastic NE and NI parts 
$F_k^{N,NE},F_k^{N,NI}$ ($N=p,n$), which correspond to processes 
$\gamma^*+N\to N$ or $\gamma^*+N\to$ (hadrons, partons). 
The NE components contribute only for $x=1$, and contain the standard  
combinations of static 
electro-magnetic form factors $G_{E,M}^N(Q^2)$ $\,\,(\eta=Q^2/(4M^2)$) 
\begin{mathletters}
\label{a3}
\begin{eqnarray}
F_1^{N,NE}(x,Q^2)&=&\frac{1
}{4}\delta(1-x)[(G_M^p)^2+(G_M^n)^2] \ ,
\label{a3a}\\
F_2^{N,NE}(x,Q^2)&=&\delta(1-x) 
\frac {[(G_E^p)^2+(G_E^n)^2+\eta \lbrace (G_M^p)^2+(G_M^n)^2\rbrace ]}
{2(1+\eta)}\ .
\label{a3b}
\end{eqnarray}
\end{mathletters}
All except $G_E^n$, have  in the past 
been assumed to be of the dipole form 
$G_d(Q^2)=[1+Q^2/0.71]^{-2}$, but recent experiments have detected
deviations  from 1 of the following quantities \cite{sill,brash,mjon}
\begin{mathletters}
\label{a4}
\begin{eqnarray}
\alpha_N &\equiv& G_M^N(Q^2)/\mu_NG_d(Q^2)\ , \qquad N=p,n\ ,
\label{a4a}\\
\gamma(Q^2) &\equiv&\frac {\mu_p G_E^p(Q^2)}{G_M^p(Q^2)}
           =\frac{G_E^p(Q^2)}{\alpha_p(Q^2) G_d(Q^2)}\ ,
\label{a4b}
\end{eqnarray}
\end{mathletters}
with $\mu_N$, the static magnetic moment of a $N$.

In the relevant $Q^2$-range, the deviation of $\alpha_p$ from 1 is moderate: 
After reaching a maximum of  $\approx 1.07$ at $Q^2\approx 2~{\rm  GeV}^2,
\,\,\,\alpha_p$  decreases and crosses the value 1 for 
$Q^2\approx 5~{\rm GeV}^2\,\,$ \cite{sill,brash}. In contrast, the
measured deviation of $\gamma$  
from 1  is far more pronounced \cite{mjon}
\begin{eqnarray}
\gamma &=1\phantom{\approx[1-0.14(Q^2-0.3)]}& 
           \qquad\qquad {\rm for\ } Q^2\lesssim 0.3\, {\rm GeV}^2\ ,
           \nonumber\\ 
       &\approx[1-0.14(Q^2-0.3)]\phantom{=1}& 
            \qquad\qquad{\rm for\ } 
            0.3 \lesssim Q^2\lesssim 5.5\, {\rm GeV}^2
\label{a99}
\end{eqnarray}
As to the NI components, for sufficiently high $Q^2$ we 
use parametrized data on $F_1^p(x,Q^2)\,$ \cite{bod} and  $F_2^p(x,Q^2)\,$ 
\cite{amad}  which are actually  averages over structures, reflecting
inclusive resonance excitations. Those stand out for relatively low $Q^2$, 
but get gradually smoothened for growing $Q^2$. For lack of direct
information on the NI parts of the SF for a neutron are frequently approximated by 
\begin{equation} 
     F_k^{n,NI}(x,Q^2)\approx F_k^{D,NI}(x,Q^2)-
     F_k^{p,NI}(x,Q^2)\ ; \qquad k=1,2\, 
\label{a5}
\end{equation}
which is reasonable for $x\lesssim 0.3$. Only recently has $F_2^n(x,Q^2)$ 
for $Q^2=3.5$ GeV$^2$ been extracted with reasonable 
accuracy \cite{rtb}. 

The above division of the $nucleon$ SF $F_k^N$ in NE and NI parts 
determines through Eq. (\ref{a2}) corresponding components $K^{A,NE},
K^{A,NI}$ in the reduced cross section defined in Eq.~(\ref{a1}). For
example,
\begin{eqnarray}
  K^{A,NE}(x,Q^2)=
  \bigg\lbrack\frac {2xM}{Q^2}F_2^{A,NE}(x,Q^2)+\frac{2}{M}F_1^{A,NE}(x,Q^2)
  {\rm tan}^2 (\theta/2)\bigg\rbrack \ ,
  \label{a7}
\end{eqnarray}
and a similar expression defines $K^{A,NI}$. Explicitly, for $I=0$ nuclei 
\cite{commar} 
\begin{mathletters}
\label{a6}
\begin{eqnarray}
  F_1^{A,NE}(x,Q^2)&=&\frac {f^{PN,A}(x)}{4}[(G_M^p)^2+(G_M^n)^2] \,
  \nonumber\\
  &=&\frac {f^{PN,A}(x)}{4} G_d^2
  [(\alpha_p \mu_p)^2+(\alpha_n \mu_n)^2]\ ,
  \label{a6a}\\
\noalign{\medskip}
  F_2^{A,NE}(x,Q^2)&=& \frac {xf^{PN,A}(x)}{2(1+\eta)}
  C_{22}(x,Q^2) \bigg [(G_E^p)^2+(G_E^n)^2+
  \eta \bigg \lbrace(G_M^p)^2+(G_M^n)^2\bigg \rbrace\bigg ]\ ,
  \nonumber\\
&=&\frac {xf^{PN,A}(x)G_d^2}{2(1+\eta)} C_{22}(x,Q^2)
  \bigg [(\gamma_c\alpha_p)^2+\eta \bigg \lbrace(\alpha_p\mu_p)^2+
  (\alpha_n\mu_n)^2 \bigg\rbrace \bigg ]\ ,
  \label{a6b}\\
\noalign{\medskip}
  \gamma_c^2&=&\gamma^2+\bigg [\frac {\mu_n\eta/\alpha_p}{1+5.6
  \eta}\bigg ]^2 \ .
\label{a6c}
\end{eqnarray}
\end{mathletters}
In. Eq. (\ref{a6c}) we have used the Galster parametrization
$G_E^n=\bigg [(\mu_n\eta G_d)/(1+5.6 \eta)\bigg ]^2\,\,$ \cite{galster} 
which approximately accounts for data for $Q^2\lesssim (1.5-2.0)\,$ GeV$^2$
~\cite{rocco}.

Using the definitions 
\begin{eqnarray}
  u(x,Q^2)&=&f^{PN,A}(x,Q^2)\alpha_p^2(Q^2)G_d^2(Q^2) \ ,
\nonumber\\
  v(x,Q^2)&=&\bigg [x^2/2(1+\eta)\bigg ] C_{22}(x,Q^2) \ ,
\label{a8}
\end{eqnarray}
one solves from  Eqs. (\ref{a7}), (\ref{a6a}) and \ref{a6b}), 
for the desired  $\alpha_n$ 
\begin{eqnarray}
  \frac {\alpha_n(Q^2)}{\alpha_p(Q^2)}=\frac{2}{\mu_n}
  \bigg [\frac {MK^{A,NE}(x,Q^2)/[2u(x,Q^2)v(x,Q^2)]- \gamma_c^2(Q^2)/4\eta}
  {1+{\rm tan}^2(\theta/2)/v(x,Q^2)} -\bigg (\frac{\mu_p}{2}\bigg )^2
  \bigg ]^{1/2} \ .
  \label{a9}
\end{eqnarray}
Should  transverse components ${\cal R}_T^{A,NE}=
F_1^{A,NE}/M$ be available, Eq. (\ref{a9}) for those reduces to
\begin{eqnarray}
  \frac {\alpha_n(Q^2)}{\alpha_p(Q^2)}=
  \frac{2}{\mu_n} \bigg [\frac{M{\cal R}_T^{A,NE}(x,Q^2)}{u(x,Q^2)}
  -\bigg (\frac{\mu_p}{2}\bigg )^2 \bigg ]^{1/2} \ .
\label{a10}
\end{eqnarray}
Next we discuss general trends of the  NE, NI components as functions 
of $x,Q^2$ in the QE region \cite{rt}. The SF $f^{PN,A}$ of a nucleus, 
composed of point nucleons, peaks around the QEP  
at $x\approx 1\, (\nu\approx Q^2/2M$), and decreases strongly with
increasing $|x-1|$. Eqs. (\ref{a6a}), (\ref{a6b}) 
then implies similar behavior of 
$F_k^{A,NE}$. As regards the variation with $Q^2$, by far the strongest ones 
are due to the static form factors $G(Q^2)$ in $F_k^{A,NE}(x,Q^2)$, which 
approximately decrease as  $\approx Q^{-4}$, while $\sigma_M$ in Eq. (\ref{a1}) at constant 
$E,\theta$ is independent of $Q^2$. 

The NI parts have entirely different characteristics. Most pronounced 
for fixed $\theta$ is their steady increase with $\nu$ (decreasing $x$), 
causing NI parts to dominate the deep inelastic region $x<1$. For 
increasing $Q^2$,  NI components decrease, but less rapid than do the NE ones. 
Ultimately NI competes with NE parts, even on the elastic side $x\ge 1$ of 
the QEP.  

The above reasoning predicts that the reduced total cross sections for 
$Q^2\ge (1.5-2.0)$ GeV$^2$ generally vary smoothly with $\nu$. 
Roughly speaking,  around the QEP, $\nu\approx Q^2/2M$, NI components 
overtake NE, which is reflected in a change of the logarithm of the 
slope of cross sections.  The above behavior has indeed been observed
for $A \ge 12$ (see Fig. 1a), for which incidentally, the normalized 
$f^{PN,A}$ hardly depend on $A\,\,$ \cite{rt}. In contrast, the non-standard 
structure of the lightest nuclei with $A\le 4$ (for instance reflected 
in the quantitatively different single-$N$ momentum distributions), 
causes the normalized $f^{PN,A}(x,Q^2)$  to be much sharper peaked, than 
is the case for $A\ge 12$.  Fig. 2 illustrates this on 
$f^{PN,A}(x,Q^2=3.0\,\, {\rm GeV}^2)$ for D, $^4$He, Fe (or C, Au), whereas 
Fig. 3 displays the $Q^2$ dependence of $f^{PN,D}(x,Q^2)$. 

From the above one predicts, that in medium and low $Q^2$ cross sections
for inclusive scattering on targets with $A\le 4$, the QEP may stand out 
against a smooth background. With increasing $Q^2$, those peaks fade into 
the NI background. Both features appear confirmed by data (cf. Fig. (1b)). 

We already argued that for decreasing $Q^2$ the NE component increases 
relative to the NI one.  Ultimately on reaches $Q_c^2\approx 
(2.0-2.5)\,$GeV$^2$, below which Eq. (\ref{a2}) is no longer  
reliable as a tool for a calculation of NI.  Yet, when wishing to extract 
information from NE parts of cross sections by their isolation, one clearly 
needs to know the relative size of the NI background. 

Another difficulty in the same $Q^2$-region regards the use of parametrized, 
resonance-averaged $F^N$, which masks actual resonance structures. In fact, 
one may exploit inclusive resonance excitation as a model for $F_k^{A,NI}$.
As is the case for the NE parts, Eqs. (\ref{a6a}), (\ref{a6b}), we expect that,
irrespective of the relatively low $Q^2$, Eq. (\ref{a2}) will properly 
produce the corresponding  $F_k^{A,res}$ due to an isolated resonance of 
moderately small width. In Appendix B we present relevant material for
$N\to\Delta$. Should the numerical outcome indeed prove that NI is 
negligibly small compared to NE, the latter can be identified with actual 
data, i.e. $K^{A,exp}\approx K^{A,NE}$.

\section{Analysis.}

In the following we shall analyze the following QE data sets:

A) Recent D-data, $E$=4.045 GeV, $\theta=15^{\circ}, 23^{\circ}\,\,$
\cite{arrd,nicu}.

B) $^4$He data for $E$=2.02 GeV, $\theta=20^{\circ}$ and $E$=
3.595 GeV, $\theta=16^{\circ}, 20^{\circ}\,\,$ \cite{day}. Those  may
well be the first QE inclusive scattering data on a nucleus, heavier than
$D$. to be used as a source for $G_M^n$.

C) Older D-data for more or less constant $Q^2=1.75, 2.50, 3.75$ GeV$^2$  
\cite{lung}, which comprise total inclusive cross sections (\ref{a9}) at
approximately the same $x,Q^2$ for various beam energies and scattering 
angles and Rosenbluth-separated transverse components. Those contain $G_M^n$ 
only in conjunction with $G_M^p$. Results for ${\cal R}_T^A$ 
have in Ref. \onlinecite{lung} been presented as effectively originating 
from data with $\theta=20^{\circ}$, which implies some binning of bands 
of $Q^2$ values. 

We start with the NI cross sections $d^2\sigma^{A,NI}$, first estimated
from inclusive $\Delta$ production (Appendix B). In Table I we both enter 
results for a $\Delta$  with its actual and a zero-width. One notices that 
the latter produces cross sections about a factor 2 lower than one with 
its actual width. This outcome warns against the use of an 
excitation amplitude into the tail of a resonance, far beyond, say, twice 
the width of the used Breit-Wigner amplitude (\ref{bp8b}). 

In addition of the above, we also performed a standard calculation
of $F^{A,NI}$ for $Q^2\lesssim 2.5 $\,GeV$^2$ using
parametrized, resonance-averaged $F_k^N$. The results are entered in the 7th
column of Table I  and enable a comparison with the resonance-excitation 
predictions. We estimate that only the entry for $Q^2\approx 2\,$ GeV$^2$ 
may be indicative of the actual size of $F^{A,NI}$.

From the results in Table I it is difficult to reach a firm conclusion 
regarding the size and $Q^2$-dependence of the  NI background
$d^2\sigma^{A,N\to\Delta}$ around the  QEP. Recalling that Eq.~(\ref{bp9}) 
give an upper limit for $d^2\sigma^{A,NI}$, we tend to conclude 
that in the QE region of the considered experiments the 
computed $\Delta$ excitation contributions are small and presumably 
negligible. Nevertheless, the conclusion is not firm, and it is desirable 
to look for corroborative evidence, which confirms  NE dominance. Only then
can one safely extract $G_M^n$  from Eq.~(\ref{a7}).

Such  support can actually  be found in a semi-empirical fashion 
directly from data, specifically on the elastic side $x\gtrsim 1$, 
$\nu\lesssim Q^2/2M$  of the QEP, and for sufficiently small $Q^2$, in
addition on its adjacent inelastic side $x\lesssim 1;\nu \gtrsim Q^2/2M$. 
In order to conclude that the data in those regions 
are essentially uncontaminated NE, and thus directly accessible to the 
extraction of $G_M^n$ by means of Eqs. (\ref{a7})--(\ref{a6b}), the 
following requirements ought to be fulfilled: 

i) In QE regions $x_{NI}(Q^2) \lesssim x\lesssim 1.1$, with $x_{NI}(Q^2)$  
the $x$-value ($<1$), where the NI part about overtakes the NE component,
the cross sections should follow  the computed bell-shaped
$x$-dependence of $f^{PN,A}(x,Q^2)$, with computed target $A$ and 
$x,Q^2$ dependence. 

ii) Extracted $\alpha_n(Q^2)$ from either Eq. (\ref{a9}) or (\ref{a10})
should not depend on the $x$-values chosen for the extraction.

iii) $\alpha_n(Q^2)$ should not depend on the target in which the neutron is
embedded.

With  $f^{PN,A}$ the source of the strongest variation with
$x$, requirements i) and ii) demand that $K^{A,NE}(x,Q^2)/u(x,Q^2)v(x,Q^2)$ 
in Eq. (\ref{a7}) be $x$-independent, and moreover, that 
tan$^2(\theta/2)/v(x,Q^2)\ll 1$. The same nuclear SF $f^{PN,A}$ carries 
the $A$-dependence, which we recall, is most pronounced for $A\le 4$: the 
ratio $K^{A,NE}/f^{PN,A}$ in Eq. (\ref{a7}) should be $A$-independent. 

The above conditions are quite stringent and lean heavily on the 
central role played by $f^{PN,A}$. Of course, it is always possible 
to fit one or two points on the elastic side of the QEP ($x\gtrsim 1$), 
whether or not the cross sections do contain some NI part in  
addition to the NE component. However, since NI parts grow with decreasing 
$x$ (increasing $\nu$), a fit of NE based on one or two points, cannot 
possibly hide a NI component over an $extended$ interval 
$x_{NI}\lesssim x\lesssim 1.1$. 

The above is most expediently tested on QE data which are represented 
on a linear scale. Figs.~4 and~5 show that criterion i) is very well met for 
recent, high-quality D data in  the elastic neighborhood of the QEP.  
As a result we could extract, for a range of selected data points, 
$\alpha_n(Q^2;x_k)$, and from those an unbiased average 
$\alpha_n(Q^2)\equiv\langle \alpha_n(Q^2)\rangle$ and an error of the mean. 

For $x$ decreasing into the inelastic region of the QEP (increasing $\nu$),
differences emerge between the measured and computed NE cross sections 
for fixed 
$\alpha_n(Q^2)$. Those reflect the growing importance of NI parts,  for
$x\lesssim x_{NI}$ and increasing with $Q^2$. 

The very quality of the fit makes one wonder why, for the stated average
$\alpha_n$, the  maxima of the two D cross sections is off by 3-5 $\%$.  
We probed, sometimes substantially larger $\alpha_n$ and the result for 
those is common to all cases to be discussed: even a 10\% increase in
$\alpha_n$ hardly  affects the NE wings and only moderately changes the 
peak area. Those bridge only a small part of the discrepancy there, while 
the error from the mean generally grows. It seems more likely that, what 
seems  to be a tiny misfit at the QEP, is actually the onset of NI at about 
the same $\nu$. In line with expectations, those are smooth in  $\nu$.

It is of course desirable to have an error estimate $\Delta\alpha_n(Q^2,x_k)$ 
due to the systematic errors in the cross sections. In spite of the fact 
that the latter are only of the order of a few $\%$, the resulting averaged 
error estimates $\langle\Delta\alpha_n(Q^2,x_k)\rangle$ may be large 
fractions of the average $\langle \alpha_n(Q^2,x_k)\rangle$. Clearly, the 
desired error estimates require far smaller systematic errors on the data 
than are presently available. The above failure actually contains 
information: provided the data are smooth and have a small error of
the mean, the method of extraction of $\alpha_n(Q^2,x_k)$ and its average is
quite sensitive to the central data. This is borne out by the above D data 
sets A).

At this point we make a digression and report on an attempt to fit the
$\theta=23^{\circ}$ D data, with $f^{PN,D}$ for $\theta=15^{\circ}$, or
alternatively with a $Q^2$-independent $f^{PN,D}$. The result, the dashed 
curve in Fig.~5, manifestly produces a far worse fit than the the drawn 
line for $f^{PN,A}$ with the $Q^2$ appropriate to $\theta=23^{\circ}$. 
The above  supports (but does not prove) the assumption that  
the SF $f^{PN,A}$ in the link (\ref{a2}) is $Q^2$-dependent, as its 
interpretation as a SF of a nucleus demands. It runs counter the claim   
that $f^{PN,A}$ is $Q^2$-independent, which holds in the PWIA (see for 
instance Ref. \onlinecite{ath}), but not for the GRS theory used above. 

Next we discuss the above mentioned older $^4$He data sets \cite{day}. As 
a comparison of Figs.~4,5 and Figs.~6--8 shows, the quality of the He data is
inferior to those for D and consequently one cannot expect a similar precision 
for $\alpha_n$, as obtained  from the above D data.

An additional complication  is the non-negligible mixing of nucleon SF in 
$F^A$,  which is primarily determined  by $C_{22}$, given by 
Eq. (\ref{eq:ap6b}). Although qualitatively understood, any evaluation 
amounts in practice to an approximation.

B1) $E=2.02\,$GeV, $\theta=20^{\circ}$: 
Fig.~6 reports our predictions for a number of $\alpha_n$. A
characteristic pattern for this case and the others mentioned below is the
insensitivity of the cross section on the elastic side for even $10\%$
changes in $\alpha_n$. 
However, those do matter around the QEP and beyond. Since the average 
$Q^2\approx 0.45$ GeV$^2$ is very low, one expects NE still to dominate 
in some range on the inelastic side of the QEP, which increases the
sample of points. From  a total of 9,  one extracts an average 
$\langle \alpha_n\rangle=1.08 \pm 0.03$. Taking out the irregular point 
$\nu=.240$ GeV close to the  QEP, the average increases to 1.10$\pm 0.03$.
Either $\alpha_n$ value is higher than most other extracted ones 
for similar $Q^2$. However, a 10\% NI contribution at the QEP and 
extrapolated behaviour about it causes an appreciable decrease of 
$\alpha_n$. With as yet no accurate NI estimate, one can only point at
sensitivity.

B2) $E=3.6\,$GeV, $\theta=16^{\circ}$:
Fig.~7 shows that, as expected, the NI component grows relative to NE
component on the inelastic side of the QEP. Limiting the sample to 9
points with $0.375$ GeV $\le \nu\le 0.495$ GeV, the average $\langle
\alpha_n\rangle=1.05 \pm 0.02$ is obtained. 

B3) $E=3.6\,$GeV, $\theta=20^{\circ}$: 
the data show substantial noise around the QEP and in 
the near-NI region (see Fig.~8). The QEP is hardly visible for this
case. One clearly cannot well fit both the elastic slope 
and the QEP region. The average over  8 points with $\nu < 0.630$ GeV
produces $\langle\alpha\rangle=1.06 \pm 0.02$.  The curves reported in
Fig.~8 are for $\alpha_n= 1.00, 1.06, 1.12$.
 
We only briefly mention  the total cross sections and separated transverse
D data of Lung \cite{lung} (sets C)). Part of those are for medium, and part 
for larger $Q^2$: all reduced data follow the theoretical predictions, 
but only to about 
10$\%$ accuracy. We note that for all $Q^2$  the data are given only
to two decimals. Therefore, in spite of the approximately fulfilled
requirement i), insufficient accuracy hampers the drawing of  sharper 
conclusions. 

To the above one may add that the extracted results may well 
be affected by the precision of the Rosenbluth separation (cf. Figs. 55 
and Table 22 in Lung's PhD Thesis \cite{lung}). The latter appears to have
been renormalized to one nominal $\theta=20^{\circ}$, which implies some 
binning. Consequently, in spite of the fact that the Rosenbluth-separated  
${\cal R}_T$ contains a simpler form for $G_M^n$ than does the total 
cross sections, we consider the latter to be a competitive and 
fiducial tool for extraction.

Table II summarizes our results for $\alpha_n(Q^2)$. Column 1 indicates 
the targets for which total QE inclusive cross sections have been analyzed, 
whereas the same for separated transverse data are denoted by ${\cal R}_T$. 
Columns 2-5 contain the beam energies, the scattering angles,  ranges of 
the considered Bjorken $x$ on the elastic side up to, and just over the QEP, 
and the corresponding ranges of $Q^2$. The separated ${\cal R^A}_T$ are all 
for fixed $Q^2$ at the QE peak and correspond to renormalized energies $E$ 
and fixed  $\theta =20^{\circ}\,$ \cite{lung}. The 6-th column gives  
ranges of the point-nucleon nuclear SF, with in parenthesis
values at the QEP. The last column presents the values of the extracted 
$\langle\alpha_n(Q^2)\rangle $, which measures the deviation of 
$G_M^n(Q^2)/\mu_n$ from a dipole form factor. As discussed above, we only
give errors of the mean values and do not include systematic errors
in the underlying data. ${\cal R}_T^{D,NE}$ between parenthesis in the last 
column are the results of Lung \cite{lung}.  

The results in Table II and a few earlier values of $\alpha_n(Q^2)$ are
shown in Fig. 9. The values, obtained in the present analysis are seen to 
agree amongst themselves and, within the experimental accuracy with 
information from other sources. 

\section{Summary and Conclusions.}

We have analyzed QE inclusive scattering on D and $^4$He. From the general 
behavior of NE components, where a nucleon in the medium absorbs a virtual 
photon without  being excited, we concluded that one should observe an 
outstanding QEP in moderate $Q^2$ cross sections for inclusive scattering 
on the lightest targets. For non-separated cross sections, those NE 
parts contain all four static form factors, as well as $f^{PN,A}(x,Q^2)$, 
the computed SF of a nucleus composed of point-nucleons. With knowledge of 
$G_{E,M}^p$ and information on $G_E^n$, the NE component of the cross 
section is a measure for  $\alpha_n(Q^2)=G_M^n(Q^2)/\mu_nG_d(Q^2)$. 

In order to assess to what extent the experimental QE cross sections 
are well represented by the uncontaminated NE component, one has to know the 
size of the NI background, relative to NE. We first 
assumed that the dominant NI parts are generated by the excitation of 
$\Delta$  resonances. In general their contributions on the elastic side 
of the QEP are small. However, those NI estimates for the QE region
in the tail of the Breit-Wigner excitation amplitude are presumably
not sufficiently precise.

In a far more reliable, semi-empirical approach, one compares the 
$x$-dependence of the reduced cross section data in the immediate 
region of the QEP with the theoretical prediction, Eqs. 
(\ref{a7})--(\ref{a6b}) for a purely NE component. Our results:

1) The values $\alpha_n(Q^2;x_k)$, extracted from the QE part of  recent 
D data, show little variation with $x_k$ and an unbiased average 
$\alpha_n(Q^2)\equiv\langle\alpha_n(Q^2)\rangle$ produces excellent fits to 
the recent D data. As expected, deviations due to NI appear on the inelastic 
side of the QEP and grow  with $\nu$ and $Q^2$.

2) The poorer quality of the He data bars an equally clean result
for the He data. Nevertheless we could extract from those
reasonable $\alpha_n$.  The one for the lowest $Q^2$ is a standard
deviation higher than other extracted values.

3) We re-analyzed Lung's non-separated  D cross sections for similar 
$x,Q^2$, but different $E,\theta$. For increasing $Q^2$, the relative weight
of $G_E^n$ grows, but simultaneously, information on $G_E^n$ becomes 
increasingly scant. We therefore only analyzed total cross sections for the 
lowest $Q^2=1.75, 2.50$ GeV$^2$ of the above experiment. 

4) The same experiment with varied kinematics provides  ${\cal R}_T$,
in principle the simplest source of $G_M^n$ from inclusive QE scattering. 
One expects the above source and unseparated data to produce the same 
$G_M^n$. The entries in Table II bear this out for $Q^2=1.75$ GeV\,$^2$, 
while Lung's value from ${\cal R}_T^{NE}$ for $Q^2=2.5$ GeV\,$^2$ somewhat 
exceeds our result. However, for the larger measured $Q^2$, our analysis 
seems to show a stronger downward trend of $\alpha_n(Q^2)$ for growing $Q^2$ 
than  reported by Lung.

It is clear from our  analysis that the extracted $\alpha_n(Q^2)$ 
are sensitive to the precision of the input. For instance, a 5$\%$ changes 
in cross sections may produce ten times larger relative changes in 
$\alpha_n(Q^2)$. The same prevents the allocation of  systematic $'$errors$'$ 
to extracted $\alpha_n$,  due to the same in the data.

We conclude that medium $Q^2$ QE inclusive scattering on light nuclei 
provide an accurate tool to determine $G_M^n$, with as   single
most important source of lack of accuracy, the same in the underlying data.

Until recently we were rather 
pessimistic as to prospects for new information. It appears however, 
that new JLab data on $^3$He have already been taken, D data are 
forthcoming, while experiments on $^4$He have been approved. Once 
analyzed, those data will be directly accessible to the above analysis 
and promise to sharpen the predictions in this communication, in 
particular for $^4$He. 

In parallel, $D(e,e'p),D(e,e'n)$ measurements will extend reliable information 
on $\alpha_n$ over a wider $Q^2$ range \cite{bro}. This will enable to 
establish whether $\alpha_p(Q^2)$ and $\alpha_n(Q^2)$ continue to behave
similarly as function of $Q^2$.

\section{Acknowledgements.}

ASR has profited from discussions with several experimentalists at JLab,
in particular with Haiyan Gao, Cynthia Keppel and Doug Higinbotham and
others. ASR thanks Jian-Ping Ling for emphasizing the 
need to ascertain the role of resonance tails in the QE region. Allison 
Lung supported our re-analyses of her NE11 data and Paul Stoler helped in 
locating inclusive $\Delta$ production data. 

\appendix 
\section{  Discussion of the  mixing coefficient in the GRS theory}
The sensitivity of the extracted $G_M^n$ from inclusive scattering data, 
in particular for low $Q^2$, calls for scrutiny in the handling of tools 
for analysis. A delicate aspect of the theory  used here concerns
the mixing coefficients entering Eq.~(\ref{a2}). All treatments and 
applications we know of are based on a comparison of hadron tensors of 
the target  and of an isolated nucleon in the PWIA of the full IS
\cite{atw,sss}. Those tensors contain invariants $p\cdot q$ and $p^A\cdot
q$, with $p$, $p^A$, the 4-momenta of the struck $N$ and the target, and
are related by the single-hole spectral function $S$ of the target 
\begin{eqnarray}                                                 
  W^{A,\mu\nu}(p^A,p^A\cdot q)=\int \frac                        
  {d^4p}{(2\pi)^4}\; S(p)\; W^{N,\mu\nu}(p,p\cdot q),           
  \label{eq:ap1}                                                      
\end{eqnarray}    
Expressing the hadron tensors by use  of the invariant SF
$F_k$, one obtains
\begin{equation}
  F_{k,0}^A(x,Q^2) = 
   \int \frac {d^4p}{(2\pi)^4}\; S(p)\; \sum_{\ell=1,2} {\cal
   C}_{k\ell}(p,\nu,|\bmq|) 
     F_k^N(\tilde x,Q^2),
   \label{eq:ap1b}
\end{equation}
with~\cite{atw,sss}
\begin{equation}
  \tilde x= {Q^2\over 2M\tilde\nu}\ ,\qquad
  \tilde\nu=\nu'+{p^2-M^2\over 2M}\ ,\qquad 
  M \nu'=p_0\nu-p_z|\bmq|\ ,\label{eq:ap1c}
\end{equation}
where $p_z$ is the component of the 3-momentum $\bmp$  of the struck
nucleon along $\bmq$. The dominant coefficient reads 
\begin{eqnarray}
  {\cal C}_{22}(p,\nu,|\bmq|)={(\nu')^2\over\nu\tilde\nu}
  \left( \bigg [1+\frac{Q^2}{|\bmq|\nu'}\frac
  {p_z}{M}\bigg ]^2 
  +\frac {Q^2}{|\bmq|^2} \bigg [\frac{\nu}{\nu'}\bigg ]^2
  \frac {[p_{_{\perp}}]^2}{2M^2}\right) \ ,
  \label{eq:ap2}
\end{eqnarray}
where $p_{_{\perp}}^2=|\bmp|^2- p_z^2$. 
The mixing coefficients ${\cal C}_{11}=1$ 
and ${\cal C}_{21}=0$ \cite{atw,sss}, while ${\cal C}_{12}$ is
negligibly small. 

We evaluate the $p_0$ integral in Eq.~(\ref{eq:ap2}) making the standard
assumption that the spectator nucleus is on its mass shell.
Energy conservation in the vertex $(A,A-1_n,N)$ then determines $p_0$.
In the target rest frame
\begin{eqnarray}
  p_0&=&M_A-\sqrt{|{\bmp}|^2+[M^n_{A-1}]^2}
  \nonumber\\
  &\approx &M_A-M^n_{A-1}-|{\bmp}|^2/2M^n_{A-1}\ ,
  \label{eq:ap3}
\end{eqnarray}
where $M_{A-1}^n$ is the mass of the $A-1$ system in the $n$-excited
state and $M_A$ the mass of the target in its ground state. In the
following, we will neglect the recoil energy of the spectator and
therefore
\begin{equation}
  p_0\approx M-E -\Delta\ ,
  \label{eq:ap3b}
\end{equation}
where $E$ the excitation  energy of the $(A-1)$ sytem
and $\Delta$ the smallest separation energy of the
$(A-1)$  nucleon system from the target. One can easily
transform the integration over $p_0$ in an integral over $E$.

We now specifically turn to GRS theory. First, whereas $f^{PN,A}$, the SF of
a nucleus composed of point-nucleons, has FSI contributions due to
scattering of off--mass shell nucleons, GRS assumes $F^N_k(\tilde
x,Q^2)$ to be the SF of an on-shell nucleon. Consequently the argument
of the nucleon SF becomes
\begin{equation}
  \tilde x\rightarrow Q^2/2M\nu'= x'  \ ,\qquad
  \tilde\nu\rightarrow\nu'\ ,
  \label{eq:ap3c}
\end{equation}
and the mixing coefficient in Eq.~(\ref{eq:ap2}) now reads
\begin{eqnarray}
 {\cal C}_{22}(p,\nu,|\bmq|)={\nu'\over\nu}\left(
 \bigg [1+\frac{Q^2}{|\bmq|\nu'}\frac
 {p_z}{M}\bigg ]^2 
 +\frac {Q^2}{|\bmq|^2} \bigg [\frac{\nu}{\nu'}\bigg ]^2
 \frac {[p_{_{\perp}}]^2}{2M^2}\right) \ .
 \label{eq:ap2b}
\end{eqnarray}
We write the GRS SF of a nucleus composed by point-nucleons  
as a lowest order term, supplemented by a
FSI  term,
\begin{equation}
   f^{PN,A}(x,Q^2)=f_0^{PN,A}(x,Q^2) + f_{FSI}^{PN,A}(x,Q^2)\ .
   \label{eq:ap4}
\end{equation}
The lowest order $f_0^{PN,A}(x,Q^2)$ can be
derived from Eq.~(\ref{eq:ap1b}) using the assumption~(\ref{eq:ap3c}).
Writing $S(p)=2\pi P(|\bmp|,E)$, the $F_2^A$ in Eq.~(\ref{eq:ap1b}) becomes

\begin{eqnarray}
  F_{2,0}^A(x,Q^2) &\approx& 
   \int \frac {d^3p}{(2\pi)^3} dE \; P(|\bmp|,E)\;  {\cal
   C}_{22}(p,\nu,|\bmq|) F_2^N(x',Q^2)\ , \nonumber\\
  &=& \int dz\;  F_2^N\bigg ({x\over z},Q^2 \bigg )
  \int \frac {d^3p}{(2\pi)^3} dE \; P(|\bmp|,E)\;  {\cal
   C}_{22}(p,\nu,|\bmq|) \delta\bigg (z-{x\over x'}\bigg )\ .
  \label{eq:ap5}
\end{eqnarray}
Introducing the Gurvitz scaling variable $y_G$~\cite{gr}
\begin{equation}
  z-{x\over x'}={|\bmq|\over M\nu}\bigg( p_z+{\nu E\over
 |\bmq|}-y_G(z)\bigg)\ , 
 \qquad y_G(z)={M\nu\over |\bmq|}\bigg(1-z-{\Delta\over M}\bigg)\ ,
 \label{eq:ap5b} 
\end{equation}
then,
\begin{eqnarray}
  F_{2,0}^A(x,Q^2) &\approx& 
  \int dz  \; F_2^N\bigg({x\over z},Q^2\bigg) \times \nonumber \\
  && \left[ {M \nu \over |\bmq|} 
  \int \frac {d^3p}{(2\pi)^3} dE \; P(|\bmp|,E)\;  {\cal
   C}_{22}(p,\nu,|\bmq|) \delta\bigg(p_z+{\nu E\over |\bmq|}-y_G(z)\bigg)
   \right]\ .
  \label{eq:ap5c}
\end{eqnarray}
The lowest order part $f_0^{PN,A}$  of the point-nucleon nuclear SF 
is defined by the expression given above between
square parenthesis. Note that (except for the factor
$M\nu/|\bmq|\equiv|dy_G(z)/dz|$)  it coincides with the expression
given in Eq.~(66) of Ref.~\cite{gr1} when ${\cal C}_{22}=1$.

Finally, the function $f^{PN,A}$ and the coefficient $C_{22}$ used in
Eq.~(\ref{a2}) are defined as 
\begin{equation}
  f^{PN,A}(z,Q^2) ={M\nu\over |\bmq|} \left[
  \int \frac {d^3p}{(2\pi)^3} dE \; P(|\bmp|,E)\;
   \delta\bigg(p_z+{\nu E\over |\bmq|}-y_G(z)\bigg)\right]+
   f_{FSI}^{PN,A}(z,Q^2) \ , 
   \label{eq:ap6}
\end{equation}
and
\begin{eqnarray}
   \lefteqn{C_{22}(z,Q^2) f^{PN,A}(z,Q^2) =}&&\nonumber\\
\noalign{\medskip}
  &&={M\nu\over|\bmq|} \left[
  \int \frac {d^3p}{(2\pi)^3} dE \; P(|\bmp|,E)
   \;  {\cal   C}_{22}(p,\nu,|\bmq|)
   \delta\bigg(p_z+{\nu E\over |\bmq|}-y_G(z)\bigg)\right]+
   \nonumber\\
\noalign{\medskip}
   && + f_{FSI}^{PN,A}(z,Q^2) \ .
   \label{eq:ap6b}
\end{eqnarray}
The expression for $f_{FSI}^{PN,A}(z,Q^2)$  can be found in
Refs.~\cite{rt,viv} and is assumed not to
be modified by  ${\cal C}$.

\section {$N \to \Delta$ inclusive cross section}

In the following we discuss the NI background in the QE region, as
due to inclusive electro-excitation of the lowest $\Delta$-resonance. Its
cross section for a proton is
\begin{mathletters}
\begin{eqnarray}
\label{bp1}
d^2\sigma^{p,NI}&\to& d^2\sigma^{p,\Delta}\approx
\sigma_M F_2^{p,\Delta}/\nu
\label{bp1a}\\
F_2^{p,\Delta}(x,Q^2)&=&\frac {Q^2}{x} {\cal N}(\Gamma_\Delta)
{\cal G}_{p\Delta}^2(Q^2)
\frac{ M_{\Delta}\Gamma_{\Delta}/\pi}
{[Q^2(1/x-1)-(M_{\Delta}^2-M^2)]^2+[M_{\Delta}\Gamma_{\Delta}]^2}
\label{bp1b}
\end{eqnarray}
\end{mathletters}
Since all data are for forward angles, it suffices to consider only $F_2$.
${\cal G}_{p\Delta}$ denotes a transition form factor to be given below and
the number ${\cal N}(\Gamma_\Delta)$ in Eq. (\ref{bp1b}) accounts for a proper
normalization of the nearly-elastic resonance amplitude.

Total cross section data are frequently  expressed in terms of those for
transverse and longitudinal virtual photons (see for instance Ref.
\onlinecite{muld})
\begin{mathletters}
\begin{eqnarray}
\label{bp2}
d^2\sigma&=&\gamma_t(\sigma_t+\epsilon\sigma_l) 
\label{bp2a}\\
\epsilon^{-1}(E;\nu,Q^2)&=&1+2\frac{{\qq}^2}{Q^2}{\rm tan}^2(\theta/2),
\label{bp2b}
\end{eqnarray}
\end{mathletters}
with
\begin{mathletters}
\begin{eqnarray}
\label{bp3}
\gamma_t(E;\nu,Q^2)&=&\sigma_M(E;\nu,Q^2)
\frac{Q^2}{4\pi^2\alpha{\qq}\epsilon(E;\nu,Q^2)},
\label{bp3a}\\  
&\approx&\frac{\alpha}{\pi^2}\frac{(E-\nu)^2}{{\qq}Q^2} \frac{1}{\epsilon}
\label{bp3b}
\end{eqnarray}  
\end{mathletters}
the flux of virtual photons. For small $\theta$ one approximates 
$\epsilon\approx\epsilon^{-1}\approx 1$, to be used in Eq. (\ref{bp3b}).

As regards  the transition form factor in  Eq. (\ref{bp1b}), we assume it 
to be of the form (cf. Eqs. (\ref{a6a}), ({\ref{a6b})) for NE).
\begin{eqnarray}
{\cal G}_{p\Delta}(Q^2)&=&\mu_{p\Delta}G_{p\Delta}(Q^2) 
\nonumber\\
G_{p\Delta}(Q^2)&=&\bigg [\frac{1}{1+Q^2/Q^2_{p\Delta}}\bigg ]^2,
\label{bp4}
\end{eqnarray}
with $\mu_{p\Delta}$, some effective transition magnetic moment and the
reduced transition form factor $G_{p\Delta}$ of a dipole form.
The  parameters in Eq. (\ref{bp4}) are estimated by a comparison of small
$\theta$ data for reduced cross sections  with Eq. (\ref{bp1b})
$$\Sigma^{p,\Delta}=d^2\sigma^{p,\Delta}/\gamma_t$$
In particular
at the top of the resonance
\begin{eqnarray}
\Sigma^{p,\Delta,{\rm max}}\approx [\sigma_M F_2^{p,\Delta,{\rm max}}
/\nu]/\gamma_t
\approx 8\pi\alpha\frac{{\qq}}{Q^2}\frac {M}{M_{\Delta}\Gamma_{\Delta}}
[\mu_{p\Delta}G_{p\Delta}(Q^2)]^2 \epsilon
\label{bp5}
\end{eqnarray}  
From data for $Q^2$=0.5, 1.0, 2.0 GeV$^2$ (Figs. 12, 13, 14 in
Ref.\cite{brasse}), we  extracted $Q^2_{p\Delta}\approx 2.7\,$GeV$^2,
\mu_{p\Delta}^2\approx 0.9$. Those values have been used in Eq. (\ref{bp4})
for all relevant $Q^2$.

No such information exists for the neutron. However, guided by the behavior
of the nucleon SF, averaged over resonances,              
$\langle F_2^p\rangle, \langle F_2^n\rangle$ (see for instance
Ref. \cite {nicu2}), it is reasonable to assume that 
\begin{eqnarray}
\bigg (F_2^{p\Delta}+F_2^{n\Delta}\bigg )/2\lesssim F_2^{p\Delta}   
\label{bp6}
\end{eqnarray} 
The above suffices to compute $F_2^{A\Delta}$ from Eq. (\ref{a2})
\begin{mathletters}
\begin{eqnarray}
\label{bp8}
  F_2^{A,\Delta}(x,Q^2)&=& \frac{Q^2}{x} [\mu_{p\Delta}G_{p\Delta}(Q^2)]^2
  I^{A,\Delta}(x,Q^2;\Gamma_\Delta)
  \label{bp8a}\\
  I^{A,\Delta}(x,Q^2;\Gamma_\Delta)&=&{\cal N}(\Gamma_\Delta)
  \bigg (\frac {M_{\Delta}\Gamma_{\Delta}}{\pi}\bigg ) \nonumber \\
  && \int_x^A dz \frac {z f^{PN,A}(z,Q^2) C_{22}(z,Q^2)}
  {[Q^2(z/x-1)-(M^2_{\Delta}-M^2)]^2+ (M_{\Delta}\Gamma_{\Delta})^2} 
\label{bp8b}
\end{eqnarray}
\end{mathletters}
Finally, the corresponding nuclear QE inclusive $\Delta$
excitation cross section reads
\begin{mathletters}
\begin{eqnarray}
\label{bp9}
\frac {d^2\sigma^{A,\Delta}(E;\theta,\nu)/A}{d\Omega d\nu}
&\lesssim& (2M)\sigma_M(E;\theta,\nu)
[\mu_{p\Delta}G_{p\Delta}(Q^2)]^2 I^{A,\Delta}(x,Q^2;\Gamma_\Delta)
\label{bp9a}\\
&\approx& \frac {2Mx}{Q^2} \sigma_M(E;\theta,\nu)
[\mu_{p\Delta}G_{p\Delta}(Q^2)]^2 (x/x_{\Delta})f^{PN,A}(x/x_{\Delta},Q^2)
\label{bp9b}\\
x_{\Delta}(Q^2)&=&\bigg [1+\frac {M_{\Delta}^2-M^2}{Q^2}\bigg ]^{-1}
\label{bp9c}
\end{eqnarray}
\end{mathletters}
with $x_{\Delta}(Q^2)$ the value of the Bjorken variable at the 
resonance peak. Eq. (\ref{bp9b}) is the zero-width limit of (\ref{bp9a}), 
which resembles the NE part, if $M_R\to M$, and thus $x_\Delta(Q^2)\to 1$. 
The same limit of $x_\Delta$ is obtained for $Q^2\to\infty$, 
corresponding to the resonance 
position in $x=1$, ultimately coinciding with the QEP.

For small, medium $Q^2$, $1/x_{\Delta}$ is substantially larger 
than 1, i.e. the  resonance peak is far from the  QE region. 
In that case, the QEP region $x\approx 1$ corresponds to 
the tail of $f^{PN,A}$, 
far from its maximum value $f^{PN,A}(x\approx 1,Q^2)$, and
consequently $d^2\sigma^{A,\Delta}$ is expected to be small.
For increasing values of $Q^2$, however, the resonance peak moves
closer and closer to the QEP and the NI contribution to the total 
cross section {\it at the QEP} can become  there quite sizable.

\begin{figure}[p]
\includegraphics[bb=-150 250 567 400,scale=.8,angle=-90]{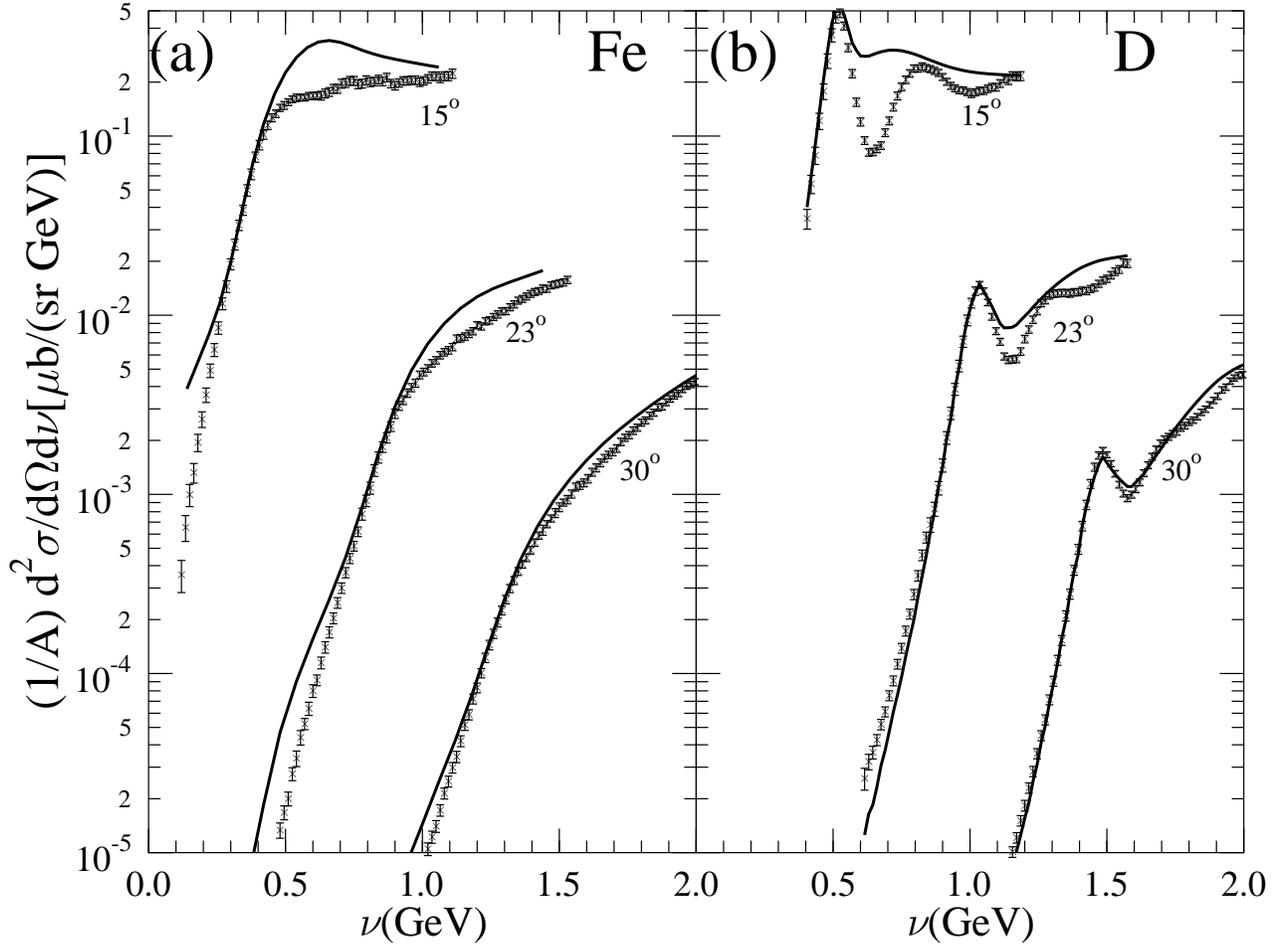}
\caption{(a) Data \protect\cite{arr2} and calculated  
\protect\cite{rt} QEP cross 
sections for inclusive scattering of $E=4.045$ GeV electrons on Fe through
$\theta=15^{\circ},23^{\circ}$ and $30^\circ$. (b)
Same as in (a) for D; data are from \protect\cite{arrd,rtd}}
\end{figure}

\begin{figure}[p]
\includegraphics[scale=.9]{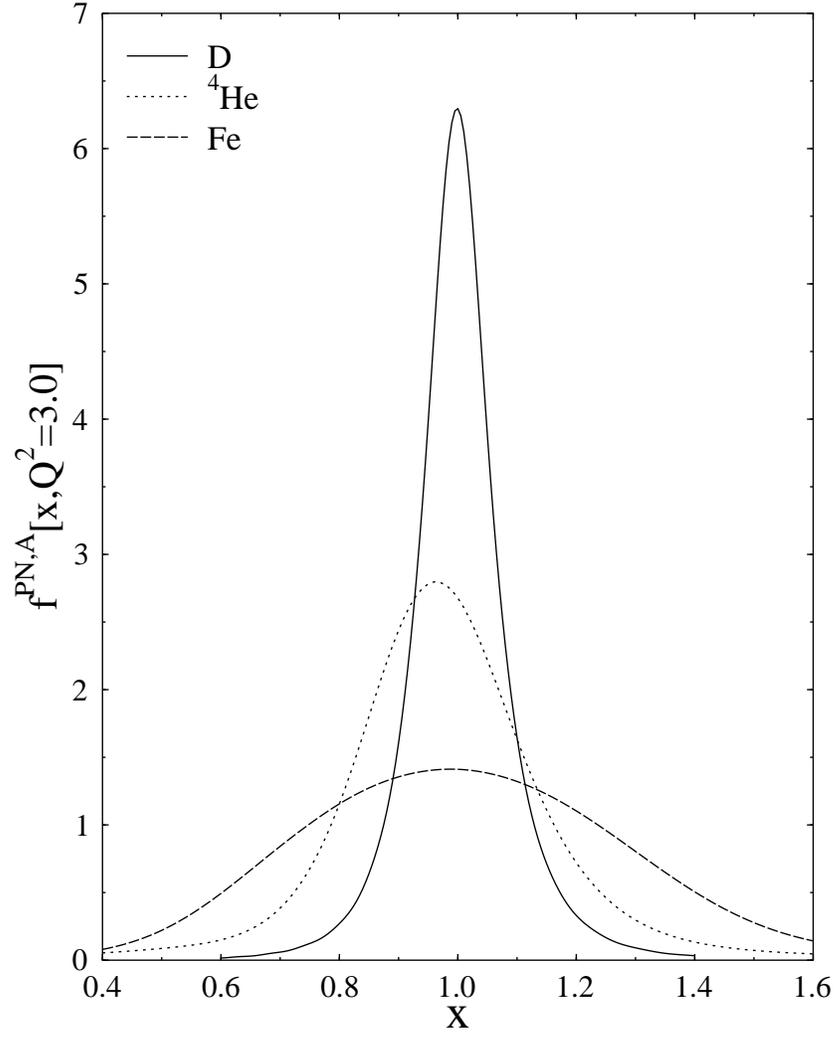}
\caption{Comparison of $f^{PN,A}(x,Q^2=3.0\, {\rm GeV}^2)$ 
for D, $^4$He, Fe.}
\end{figure}
 
\begin{figure}[p]
\includegraphics[scale=.9]{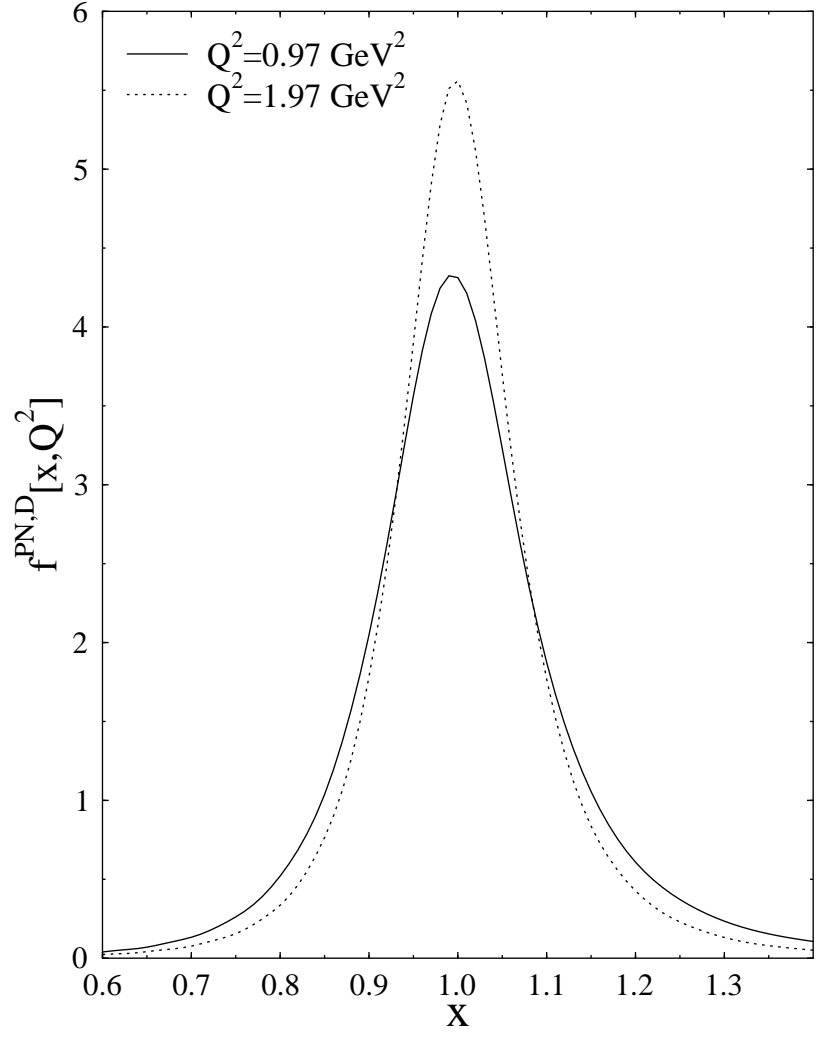}
\caption{The SF $f^{PN,D}(x,Q^2)$ for $Q^2=0.972, 1.94\,$GeV$^2$.}
\end{figure}

\begin{figure}[p]
\includegraphics[bb=-150 450 567 400,scale=.7,angle=-90]{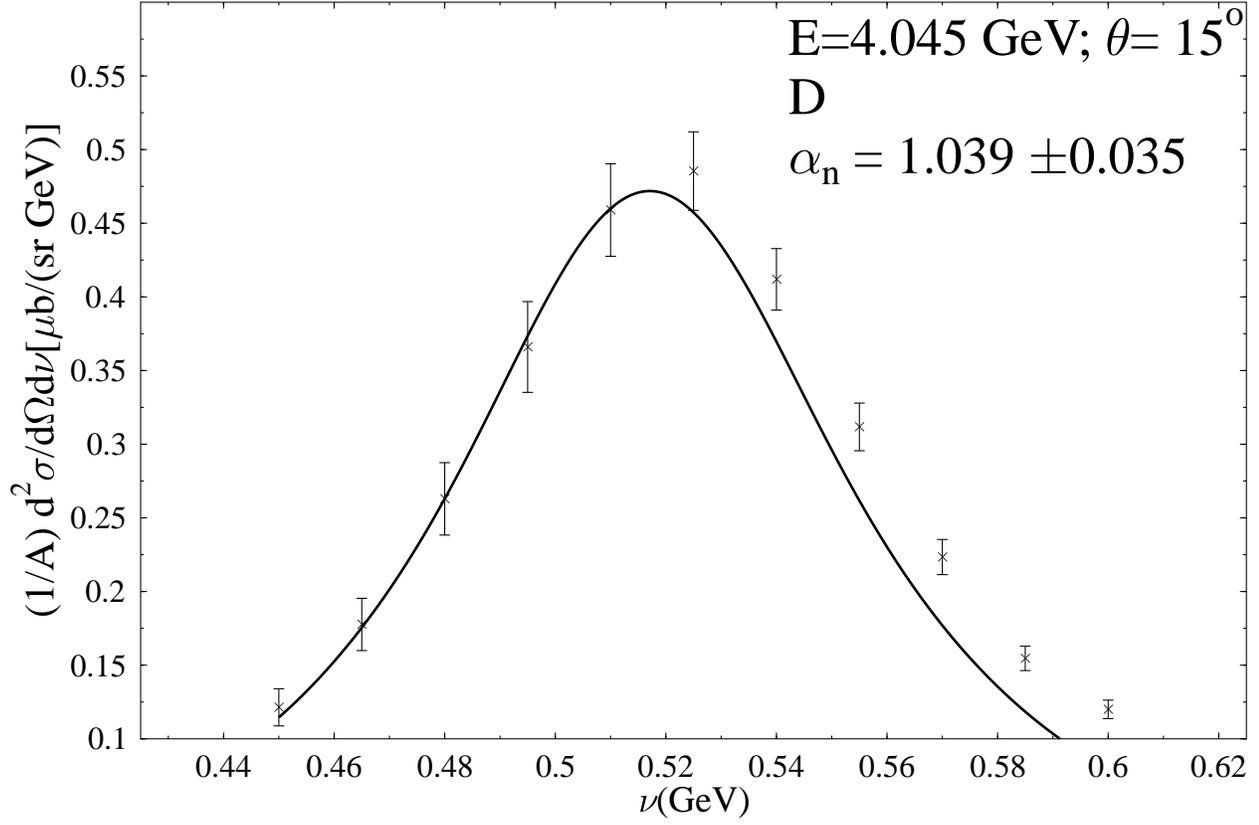}
\caption{Cross section for QE inclusive scattering of $E=4.045$ GeV 
electrons on D for $\theta=15^{\circ}$.
The drawn line is the theoretical NE cross section for
$\alpha_n=1.039$.}
\end{figure}

\begin{figure}[p]
\includegraphics[bb=-150 450 567 400,scale=.7,angle=-90]{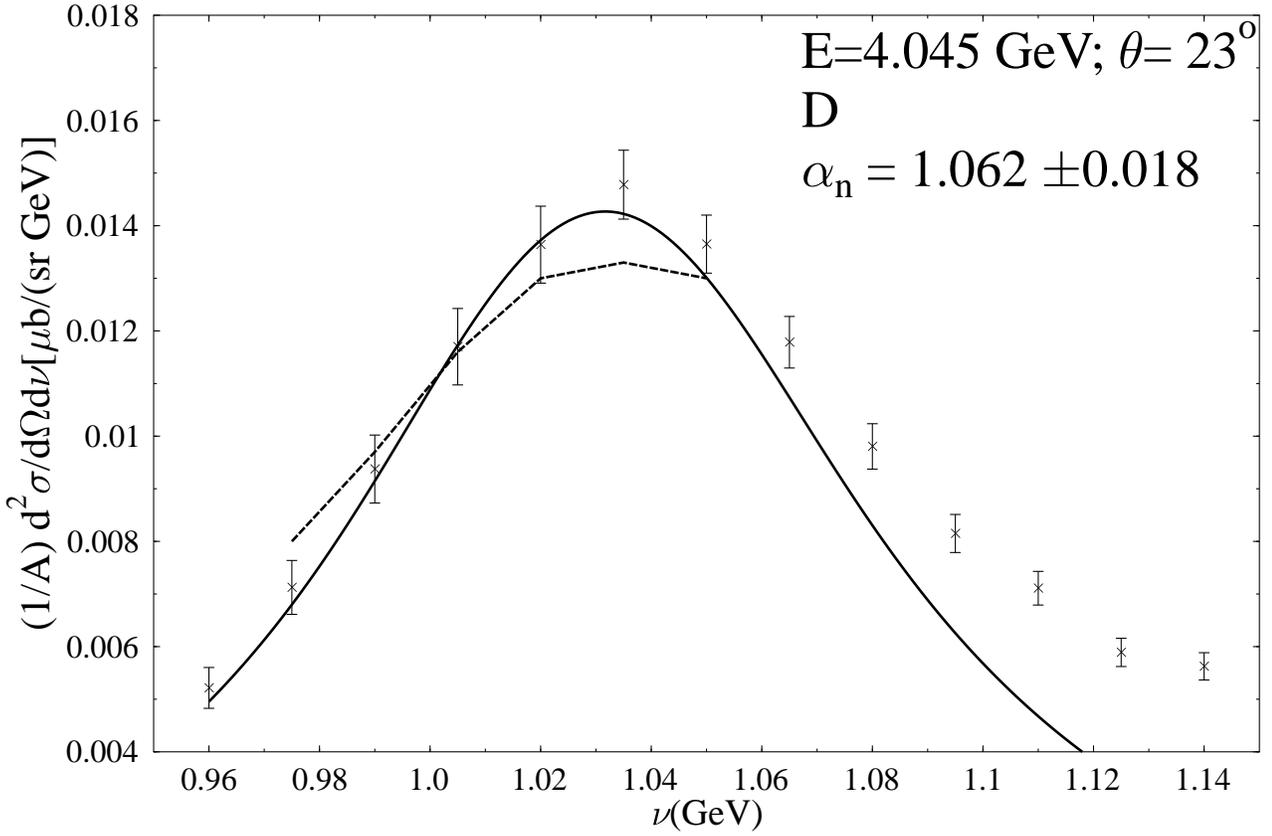}
\caption{The same as in Fig. 4 for $\theta=23^{\circ}$. 
The drawn line is the theoretical NE cross section for
$Q^2=1.94\,$GeV$^2$ and $\alpha_n=1.062$.
The dotted line represents the result of a calculation with the SF 
$f^{PN,D}(x,Q^2=0.972\,$GeV$^2$), 
instead of the same with the value
$Q^2=1.94\,$ GeV$^2$, pertinent to this case $\theta=23^{\circ}$.}
\end{figure}

\begin{figure}[p]
\includegraphics[scale=.75]{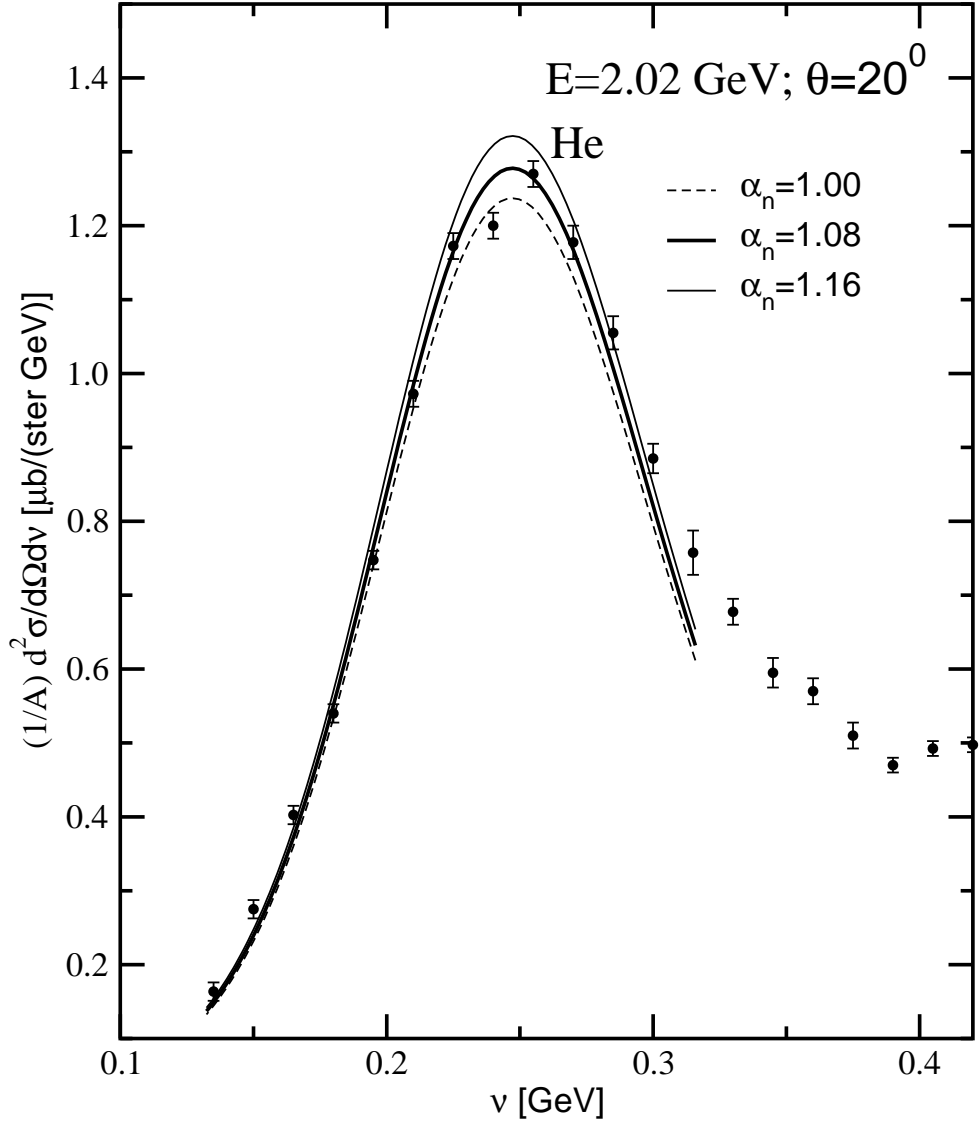}
\caption{Cross section for QE inclusive scattering of $E=2.02$ GeV 
electrons on He for $\theta=20^{\circ}$. Data are 
from Ref.~ \protect\cite{day}.
The lines are the theoretical NE cross sections for
three values of $\alpha_n(Q^2)$.
The unbiased average value of $\alpha_n(Q^2)$ for this case
can be found in Table II.}
\end{figure}

\begin{figure}[p]
\includegraphics[scale=.75]{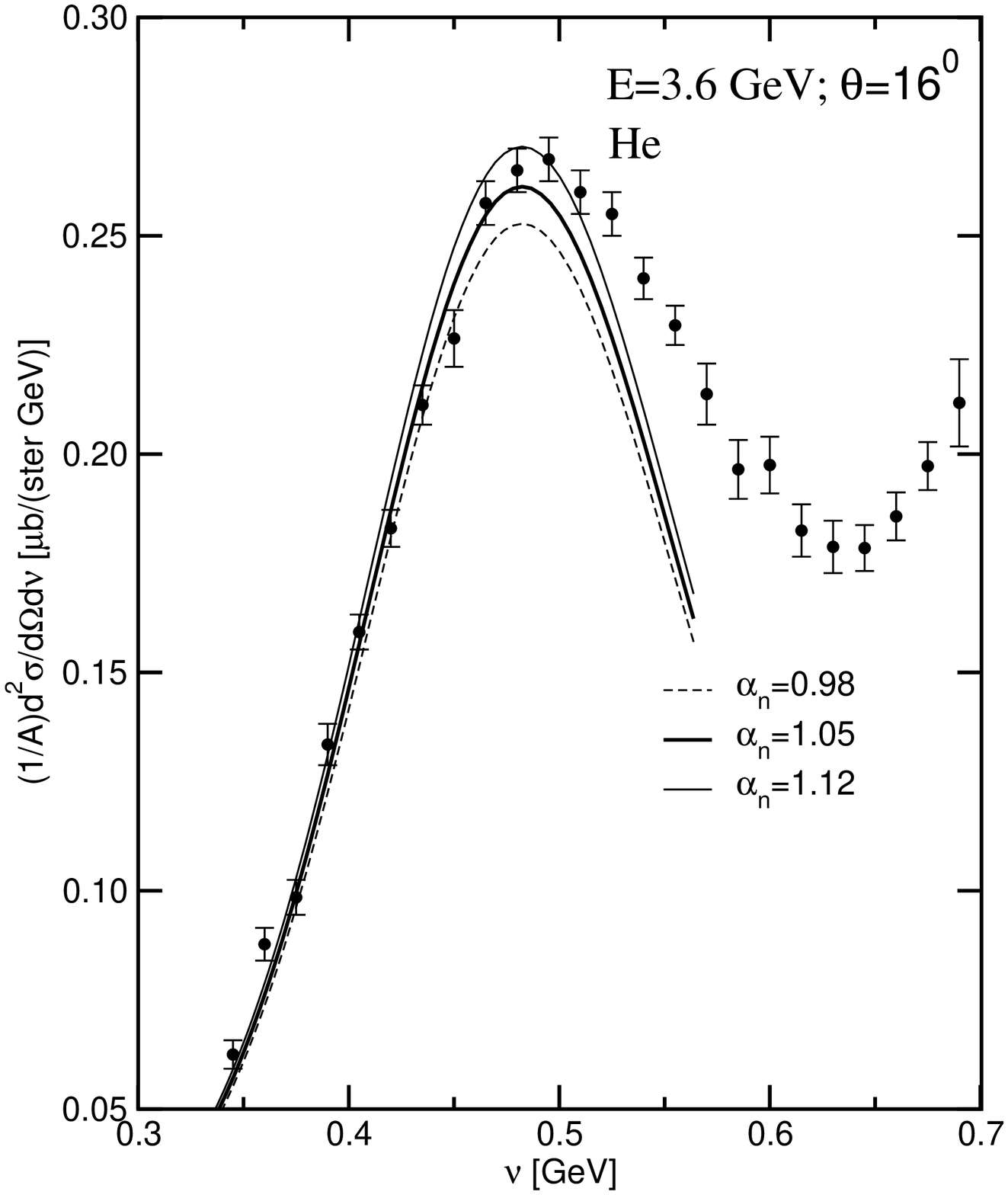}
\caption{The same as in Fig.~6 for
$E$=3.595 GeV, $\theta=16^{\circ}$.}
\end{figure}

\begin{figure}[p]
\includegraphics[scale=.75]{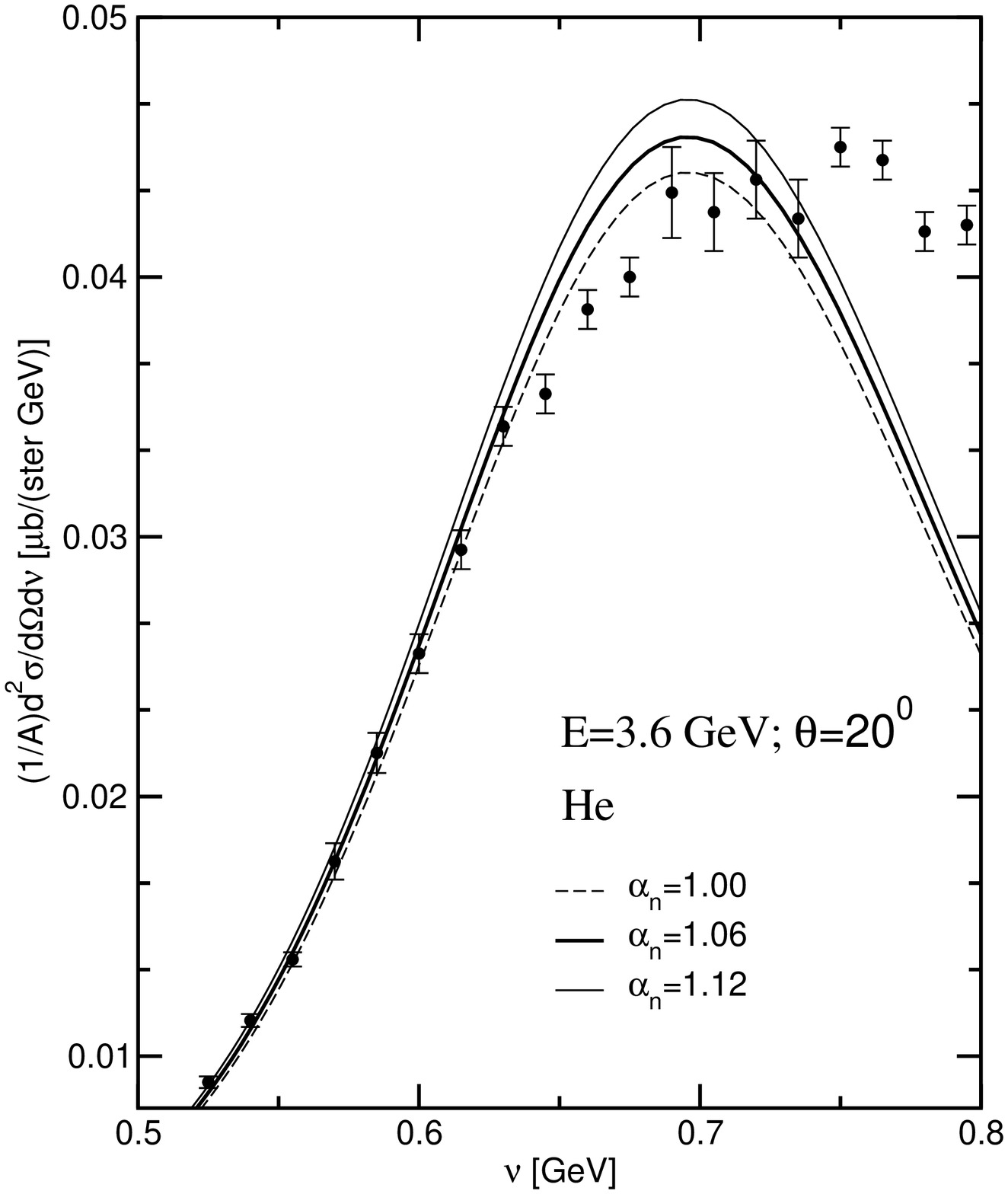}
\caption{The same as in Fig.~6 for
$E$=3.595 GeV, $\theta=20^{\circ}$.}
\end{figure}

\begin{figure}[p]
\includegraphics[bb=-150 440 567 400,angle=-90,scale=.65]{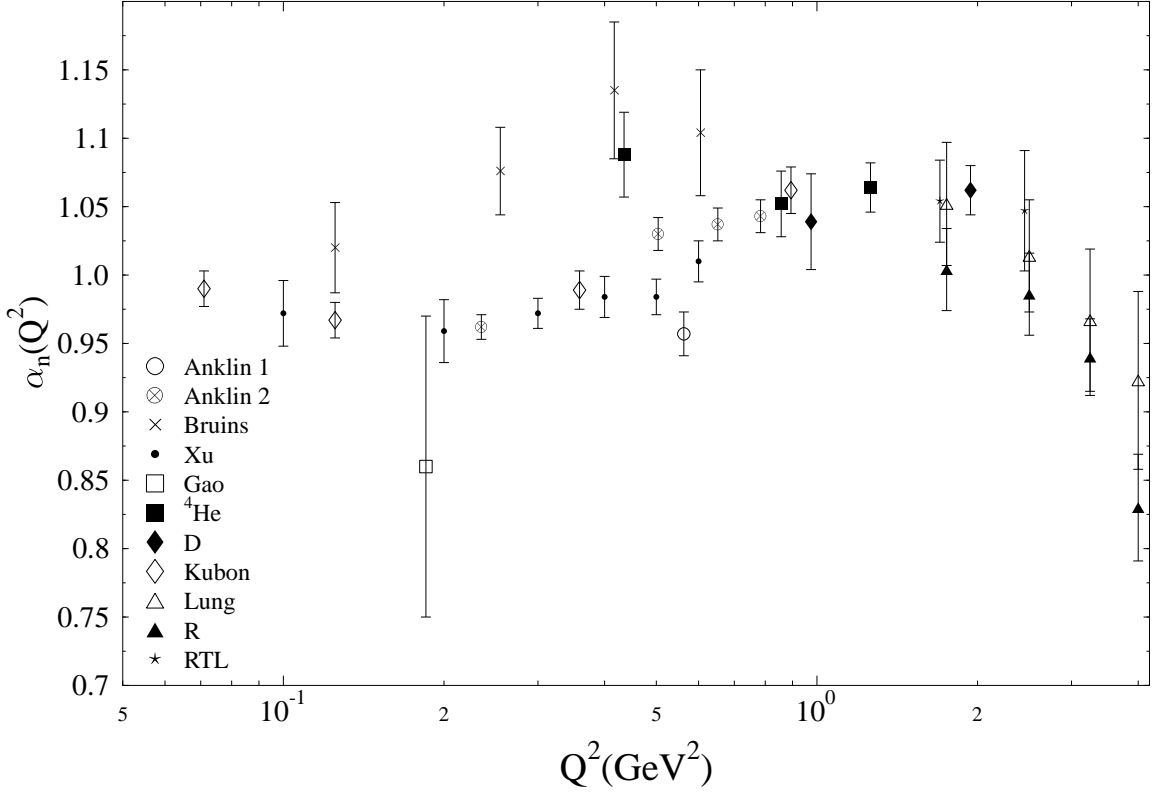}
\vspace{1cm}
\caption{$\alpha_n=G_M^n/\mu_n G_d$ as function of $Q^2$. Entered are
previous results and those obtained in the present work 
(filled squares, diamonds and triangles).}
\end{figure}


\begin{table}[p]
\caption {\protect
$N\to\Delta\,$ NI inclusive cross sections for D, $^4$He. 
Columns 1--4 give target, beam energy and scattering angle,
$\langle Q^2\rangle$ and $x$-position of the resonance for a number of 
values of the energy loss $\nu$ and $x$ around the QEP. 
Moreover, in columns 5, 6 we report the 
NI cross sections computed
with the $N\to \Delta$ excitation model described in Appendix~B
using $\Gamma_\Delta=0.12$ GeV and $\Gamma_\Delta=0$, respectively.
In column 7, we  report the NI cross section computed with
Eq.~(\ref{a2}) using the parametrized, resonance-averaged, nucleon 
SF $F^{N,NI}_k(x,Q^2)$. Finally, in the last column we report the 
measured (total) inclusive cross sections. All quantities are in powers of 
GeV; cross sections are in $\mu$b/ster/GeV}.
\begin{tabular}{c@{$\;$}|@{$\;$}
                c@{$\;$}|@{$\;$}
                c@{$\;$}|@{$\;$}
                c@{$\;$}|@{$\;$}
                c@{$\;$}|@{$\;$}
                c@{$\;$}|@{$\;$}
                c@{$\;$}|@{$\;$}c}
\hline
target    & $E,\theta$ & $\langle Q^2\rangle,x_{\Delta}$  & $\nu,x$ &
${1\over A} d^2\sigma^{A,\Delta;\Gamma_\Delta}$ & 
${1\over A} d^2\sigma^{A,\Delta;\Gamma_\Delta=0}$ & 
${1\over A} d^2\sigma^{A,NI}$ & 
${1\over A} d^2\sigma^{A,{\rm total}}_{\rm exp} $\\
\hline
$D\,${\protect\cite{arrd,nicu}}
    & 4.045, $15^{\circ}$ & 0.972, 0.601& 0.465, 1.131 &
0.0193   &  0.0089  & 0.0162 & 0.178   \\
     &      &                   & 0.495, 1.054 &   
0.0368   &  0.0130  & 0.0225 & 0.263   \\ 
     &      &                   & 0.525, 0.985 &
0.0656   &  0.0193  & 0.0627 & 0.435   \\
     &      &                   & 0.555, 0.922 &
0.0827   &  0.0295  & 0.1110 & 0.312   \\
\hline
 $D\,${\protect\cite{arrd,nicu}}
     & 4.045, $23^{\circ}$ & 1.94, 0.750& 0.975, 1.079 &
0.00195  &  0.00096 &0.00050 & 0.0064 \\
     &      &                  & 1.005, 1.037 &
0.00325  &  0.00150 &0.00084 & 0.0108 \\
     &      &                  & 1.035, 0.997 &
0.00531  &  0.00225 &0.00138 & 0.0248 \\
     &      &                  & 1.065, 0.959 &
0.00806  &  0.00363 &0.00232 & 0.0126 \\
\hline
$^4$He$\,${\protect\cite{day}}
      & 2.02, $20^{\circ}$ & 0.434, 0.402& 0.210, 1.125 &
0.0580   &  0.0202  & 0.256  & 0.973  \\
      &     &                         & 0.225, 1.035 &
0.0833   &  0.0281  & 0.386  &  1.173 \\
      &     &                         & 0.240, 0.962 & 
0.1122   &  0.0382  & 0.535  &  1.200 \\
      &     &                         & 0.255, 0.898 &  
0.1442   &  0.0504  & 0.704  &  1.270 \\
\hline
$^4$He$\,${\protect\cite{day}}
      & 3.595, $16^{\circ}$& 0.872, 0.575& 0.420, 1.121 &
0.0322   & 0.0165   & 0.0297  & 0.183  \\          
      &     &                         & 0.450, 1.037 & 
0.0520   & 0.0259   & 0.0488  & 0.227  \\              
      &     &                         & 0.465, 0.998 & 
0.0629   & 0.0318   & 0.0607  & 0.258  \\              
      &     &                         & 0.480, 0.963 & 
0.0824   & 0.0399   & 0.0766  & 0.227  \\              
\hline
$^4$He$\,${\protect\cite{day}}
      & 3.595, $20^{\circ}$& 1.266, 0.662& 0.615, 1.119 & 
0.008    & 0.0039   & 0.0029 & 0.0293  \\              
      &     &                          & 0.645, 1.056 & 
0.015    & 0.0086   & 0.0069 & 0.0343  \\              
      &     &                         & 0.675, 0.999 & 
0.021    & 0.0126   & 0.0105 & 0.0400  \\              
      &     &                         & 0.705, 0.947 & 
0.029    & 0.0186   & 0.0144 & 0.0425  \\              
\hline
\end{tabular}
\label{TableI}
\end{table}


\begin{table}[p]
\caption {Extraction of $\alpha_n(Q^2)$ from QE inclusive scattering 
data on D, $^4$He. Columns 1--4 give the target, the beam energy $E$, 
the scattering angle $\theta$ and the range of values of the Bjorken $x$
variable chosen to perform the extraction of $\alpha_n(Q^2)$. Column 5 gives
the corresponding range of values for $Q^2$.
Column 6 give the SF $f^{PN,A}(x,Q^2)$ for the extreme
values of $x$ in the range considered, and, in parenthesis, its
maximal values reached when $x\approx 1$.
The last column
gives $\alpha_n(Q^2)$ with error of the mean  over the considered
$x$-range. The  values between parenthesis are Lung's results without error
bars. }
\begin{tabular}{c|c|c|c|c|c|c}
\hline
target    & $E$\ [GeV] & $\theta$  & $x$ & $Q^2\ {\rm [GeV^2]}$ &
$f^{PN,A}(x,Q^2)$ & $\alpha_n(Q^2)$\\
\hline
 $^4$He$\,${\protect\cite{day}}
        & 2.02  & $20^{\circ}$ & 1.018-0.745 & 0.444-0.430
&  1.18-1.20 (1.49)  & 1.08$\pm 0.03$ \\
        & 3.595 & $16^{\circ}$ & 1.041-0.908 & 0.887-0.864
&  1.57-1.92 (1.92) & 1.05$\pm 0.02$ \\
        & 3.595 & $20^{\circ}$ & 1.126-0.905 & 1.275-1.250
&  1.28-2.11 (2.16) & 1.06$\pm 0.02$ \\
\hline
   $D\,${\protect\cite{arrd,nicu}}
        & 4.045 & $15^{\circ}$ & 1.131-0.953 & 0.988-0.972
&  1.31-3.65 (4.30)  & 1.039 $\pm 0.035$  \\
        & 4.045 & $23^{\circ}$ & 1.079-0.978 & 1.976-1.929
&  2.44-5.18 (5.18)   & 1.062 $\pm 0.018$ \\
\hline
 $D\,${\protect\cite{lung}}
        & 5.507 &  $15.2^{\circ}$ & 1.063-0.978 & 1.769-1.741
&  2.89-5.04 (5.31) & 1.055 $\pm0.047$  \\
        & 2.407 & $41.1^{\circ}$ & 1.081-0.957 & 1.803-1.721
&  2.37-4.89 (5.32)  & 1.050 $\pm0.017$  \\
        & 1.511 & $90.0^{\circ}$ & 1.059-0.977 & 1.812-1.728
&  3.21-4.79 (5.26)  & 1.057 $\pm0.023$  \\
\hline
  ${\cal R}_T^{D,NE}$ &3.809 & $20^{\circ}$ &1.141-0.962
&$<Q^2>$=1.75
&  1.79-3.38 (5.31) & 1.004$\pm0.030 \,\bigg (1.052\,^{3}\bigg )$ \\
\hline
 $D\,${\protect\cite{lung}}
      & 5.507 & $19.0^{\circ}$ & 1.104-1.000 & 2.561-2.501
&  1.69-5.65 (5.98)  & 1.032 $\pm 0.035$  \\                        
      & 2.837 & $45.0^{\circ}$ & 1.101-0.991 & 2.613-2.500
&  1.69-5.91 (5.94)  & 1.031 $\pm 0.043$  \\                        
      & 1.968 & $90.0^{\circ}$ & 1.064-0.984 & 2.608-2.474 
&  3.06-5.71 (5.90)  & 1.078 $\pm 0.055$  \\                         
\hline
  ${\cal R}_T^{D,NE}$ &5.016 & $20^{\circ}$ & 1.068-0.940& 
$\langle Q^2 \rangle $=2.50
&  2.92-4.16 (5.94)  &0.986$\pm 0.030$ \bigg (1.014$\,^{3}\bigg )$ \\
\hline
  ${\cal R}_T^{D,NE}$ &5.016 & $20^{\circ}$ & 1.051-0.958& 
$\langle Q^2 \rangle$=3.25                 
&  3.50-6.15 (6.43)  &0.940$\pm 0.028$ \bigg (0.967$\,^{3}\bigg )$ \\
\hline 
  ${\cal }{\cal R}_T^{D,NE}$ &5.016 & $20^{\circ}$ & 1.079-1.038& 
$\langle Q^2 \rangle$=4.00 
&  3.80-6.20 (6.50)  &0.830$\pm 0.040$ \bigg (0.923$\,^{3}\bigg )$ \\ 
\hline
\end{tabular}
\label{TableII}
\end{table}


\begin{references}

\bibitem{ank1} 
H. Anklin $et\, al$, Phys. Lett. B336, 313 (1994); $ibid$
B428, 248 (1998); E.E.W. Bruins $et\,al$ Phys. Rev. Lett. 75, 21 (1995).

\bibitem{kubon}
G. Kubon $et\,al$, Phys. Lett. B524, 26 (2002).

\bibitem{lung}
A. Lung $et\, al$, Phys. Rev. Lett. 70, 718 (1993); PhD. thesis, The
American University, Washington D.C., 1992.

\bibitem{gao} 
H. Gao $et\, al$, Phys. Rev. C 50, R546 (1994).

\bibitem{xu}
W. Xu $et\,al$, Phys. Rev. Lett. 85, 2900 (2000); F. Xiong $et\, al$,
Phys. Rev. Lett 87, 242501 (2002); W. Xu $et\,al$, Phys. Rev. C 67, 012201
(2003).

\bibitem{golak}
J. Golak $et\, al$, Phys. Rev. C 66, 024008 (2002) 

\bibitem{jones2}
M. Jones, private communication. 

\bibitem{bench}
H. Kamada $et\,al$, Phys. Rev. C 64, 044001 (2001).
         
\bibitem{arrd}
J. Arrington $et\,al$, Phys. Rev. C 64, 014602 (2001).

\bibitem{nicu}
I. Niculescu $et\,al$. Phys. Rev. Lett. 85, 1182 (2000).

\bibitem{day}
D.B. Day $et\, al$, Phys. Rev. C 48, 1849 (1993).

\bibitem{sill}   
A.F. Sill $et\,al$, Phys. Rev. D 48, 29 (1993); L. Andivahis $et\, al$,
Phys. Rev. D50, 5491 (1994).

\bibitem{brash}  
E.J. Brash, A. Kozlov, Sh. Li, G.M. Huber, Phys. Rev. C 65, 051001 (2002).

\bibitem{mjon}   
M. Jones $et\,al$, Phys. Rev. Lett. 84, 1398 (2000); Third Workshop on
'Perspective in Hadronic Physics' Trieste 2001, IT; to be published;
O. Gayou $et \, al$, Phys. Rev. C64, 038202 (2001).

\bibitem{rocco}  
R. Schiavilla and I. Sick, Phys. Rev. C 64, 041002(R) (2001).

\bibitem{gr}
S.A. Gurvitz and A.S. Rinat, TR-PR-93-77/ WIS-93/97/Oct-PH; Progress in
Nuclear and Particle Physics, Vol. 34, 245 (1995).

\bibitem{atw}
G.B. West, Ann. of Phys. (NY) 74, 646 (1972); W.B. Atwood and G.B. West,
Phys. Rev. D7, 773 (1973).

\bibitem{sss} 
M.M. Sargsian, S. Simula and M.I. Strikman, Phys. Rev. C66, 024001
(2002)

\bibitem{ciof}
See for instance: C. Ciofi degli Atti, E. Pace and G. Salme, Phys. Rev.
C 43, 1155 (1991).

\bibitem{grs}   
H. Gersch, L.J. Rodriguez and Phil N. Smith, Phys. Rev. A5, 1347 (1973).

\bibitem{gr1}
S.A. Gurvitz and A.S. Rinat,  Phys. Rev. C 65, 024310 (2002).

\bibitem{rt}
A.S. Rinat and M.F. Taragin, Nucl. Phys. A598, 349 (1996); $ibid$ A620,
412 (1997); Erratum: $ibid$ A623, 773 (1997); Phys Rev. C 60, 044601 (1999).

\bibitem{lle}
C.H. Llewelyn Smith, Phys. Lett. B 128 (1983) 107; M. Ericson and A.W.
Thomas, $ibid$ p. 112.

\bibitem{bod}
A. Bodek and J.L. Ritchie, Phys. Rev.  D23, 1070 (1981)

\bibitem{amad}
P.  Amadrauz $et\,al$, Phys.  Lett.  B295, 159 (1992); M.  Arneodo $et\,
al\,,ibid$ B364, 107 (1995)

\bibitem{rtb}
A.S. Rinat and M.F. Taragin, Phys. Lett. 551, 284 (2003).

\bibitem{commar}   
A.S. Rinat and M.F. Taragin, Phys. Rev. C 62, 034602 (2000).
         
\bibitem{galster}
S. Galster $et\,al$, Nucl. Phys. B32, 221 (1971).
        
\bibitem{arr2}
J. Arrington $et\,al$. Phys. Rev. Lett. 82, 2056 (1999).

\bibitem{rtd}
A.S. Rinat and M.F. Taragin, Phys. Rev. C 65, 042201(R) (2001).

\bibitem{ath}
R.L. Jaffe, Nucl. Phys. A478, 3c (1988); R.P. Bickerstaff and A.W. Thomas, 
J. Phys. G 15, 1523 (1989). 

\bibitem{viv}
M. Viviani, A. Kievsky and A.S. Rinat, nucl-th/0111049, Phys. Rev. C67,
034003 (2003).

\bibitem{gur}
S.A. Gurvitz, Phys. Rev. C 42, 2653 (1990).
         
\bibitem{muld}
P.J. Mulders, Phys. Reports, 185, 83 (1990).

\bibitem{brasse}
F.W. Brasse $et\,al$, Nucl. Phys. B 110, 413 (1976).

\bibitem{nicu2}
E.D. Bloom and F.J. Gilman, Phys. Rev. Lett. 25, 1140 (1970); Phys. Rev. D
4, 2901 (1971); I. Niculescu $et\,al$, Phys. Rev. Lett. 85, 1186 (2000).

\bibitem{bro}
W. Brooks, private communication.

\end{references}
\end{document}